\documentclass[12pt]{article}
\usepackage{graphicx}
\usepackage[cp1251]{inputenc}
\begin{document}
\pagestyle{empty}
\pagestyle{headings} {\title{The Higgs boson
decays with the lepton flavor violation}}
\author{O.M.Boyarkin\thanks{E-mail:oboyarkin@tut.by}, G.G.Boyarkina,
D.S.Vasileuskaya\\
\small{\it{Belorussian State University,}}\\
\small{\it{Dolgobrodskaya Street 23, Minsk, 220070, Belarus}}}
\date{}
\maketitle
\begin{abstract}
Within the left-right symmetric model (LRM) the decays
$$S_1\to\mu^++\tau^-,\qquad S_1\to\mu^-+\tau^+$$
where $S_1$ is an analog of the standard model Higgs boson, are
considered. The widths of this decays are found in the third order
of the perturbation theory. Since the main contribution to the
decay widths is caused by the diagram with the light and heavy
neutrinos in the virtual state then investigation of this decays
could shed light upon the neutrino sector structure.

The obtained decay widths critically depend on the charged gauge
bosons mixing angle $\xi$ and the heavy-light neutrinos mixing
angle $\varphi$. The LRM predicts the values of these angles as
functions of the vacuum expectation values $v_L$ and $v_R$. Using
the results of the existing experiments, on looking for the
additional charged gauge boson $W_2$ and on measuring the
electroweak $\rho$ parameter, gives
$$\sin\xi\leq5\times10^{-4},\qquad\sin\varphi\leq2.3\times10^{-2}.
$$ However, even using the upper bounds on $\sin\xi$ and
$\sin\varphi$ one does not manage to get the upper experimental
bound on the branching ratio $\mbox{BR}(S_1\to\tau\mu)_{exp}$
being equal to $0.25\times10^{-2}$. The theoretical expression
proves to be on two orders of magnitude less than
$\mbox{BR}(S_1\to\tau\mu)_{exp}$.
\end{abstract}
\hspace{5mm}

\hspace*{-6mm}{\it{Keywords}}: Higgs boson, lepton flavor
violation, left-right symmetric model, heavy and light neutrinos,
mixing in the neutrino sector, Large Hadron Collider.\\[5mm]

\section{Introduction}

Upon discovering the Higgs boson, the obvious next step is to
elucidate if it is an elemental or a composite particle and if
there is physics beyond the Standard Model (SM) that could be
hidden in the Higgs sector. Expectation for departure from SM
behavior are based on the following facts. The SM has not found
satisfactory explanation of baryon asymmetry of the Universe,
neutrino mass smallness, the value of the muon anomalous magnetic
moment, hierarchy problem and so on. Moreover, among the SM
particles there are no candidates on the role of weakly
interacting massive particles which enter into the non-baryonic
cold dark matter.

It is clear that the future ambitious experimental program, both
at the upgraded Large Hadron Collider (LHC) and future linear
colliders, which will determine all the Higgs couplings with
higher precision than at present, will play a central role. A
particularly interesting possible departure from the Higgs
standard properties will be Higgs decays going with lepton flavor
violation (LFV). These decays do not take place even in the
minimally extended SM (SM with massive neutrinos), since lepton
flavor symmetry is an exact symmetry of the SM and therefore it
predicts vanishing rates for all these LFV processes to all orders
in perturbation theory. It should be noted that any experimental
signal of LFV will indicate that some new physics, either new
particles or new interactions must be responsible for it.

The ATLAS and CMS collaborations are actively searching for these
LFV Higgs decays. For example, the CMS collaboration saw an excess
on the $H\to\tau\mu$ channel after the run-I (this process
includes both $H\to \mu^+\tau^-$ and $H\to \mu^-\tau^+$), with a
significance of $2.4\sigma$ and a value \cite{VK15,Gaa}
$$\mbox{BR}(H\to\tau\mu)=(0.84^{+0.39}_{-0.37})\%.\eqno(1)$$
However, neither this excess, nor other positive LFV Higgs decay
signal, have been detected at the present run-II. As of now, ATLAS
has released their results after analyzing 20.3 $\mbox{fb}^{-1}$
of data at a center of mass energy of $\sqrt{s}=8$ TeV, achieving
sensitivities of the order of $10^{-2}$ for the $H\to\tau\mu$ and
$H\to\tau e$ channels \cite{GA17}. CMS has also searched for the
$H\to\mu e$ channel after the run-I \cite{VKH16} and has further
enhanced the sensitivities of the $H\to\tau\mu$ and $H\to\tau e$
channels with new run-II data \cite{CMS17} of $\sqrt{s}=13$ TeV,
setting the most stringent upper bounds for the LFV Higgs decays,
that at the 95\% CL are as follows
$$\mbox{BR}(H\to\mu e)<3.5\times10^{-4}\eqno(2)$$
$$\mbox{BR}(H\to\tau e)<0.61\times10^{-2}\eqno(3)$$
$$\mbox{BR}(H\to\tau\mu)<0.25\times10^{-2}\eqno(4)$$

There is no question that observation of the Higgs boson decay
with the LFV is a smoking gun signal for physics beyond the SM.
These decays have been studied for a long time in the literature
within various SM extensions (for recent works see,
\cite{KC16,SB16,Eah16,AA16}).

The models predicting the Higgs boson decays with LFV could be
classified into two categories. Among the first are the SM
extensions in which existence of these decays is provided by
introducing the Higgs boson LFV couplings by hand. This can be
achieved by an extension of the scalar sector with some additional
discrete symmetries (see, for example, Ref. \cite{MDCa,ACri}). It
is clear that all these SM extensions necessarily introduce a
number of new arbitrary parameters. Notice that in the models of
this kind the Higgs decays (2)-(4) proves to be allowed even at
the tree approximation.

However, the more elegant explanation of the Higgs decays with LFV
gives models falling into the second category in which the flavor
mixing among particles of different generations is embedded by the
construction. Example is provided by the supersymmetric models in
which the flavor mixing among the three generations of the charged
sleptons and/or sneutrinos takes place. This mixing produces via
their contributions the Higgs decay channel $H\to
l_i\overline{l}_j$ at the one-loop level \cite{JLD00,AAD14}.
Another example is the left-right symmetric model (LRM)
\cite{ICP74,RNM75,GSRN}, where the LFV processes are caused by the
mixing in the neutrino sector. Within the LRM the LFV was
investigated by the example of the processes \cite{Bom97}
$$e^-+\mu^+\to W_k^-+W_n^+,\qquad e^-+\mu^-\to W_k^-+W_n^-,$$
which may be observed on the muon colliders and the decays
\cite{Bom04}
$$\mu^-\to e^++e^-+e^-,\qquad \mu^-\to e^-+\gamma.$$
In so doing one was shown that within the LRM it could be possible
to obtain the upper experimental bounds on the BR($\mu^-\to
e^+e^-e^-)$ and BR($\mu^-\to e^-\gamma).$ In this work we also
investigate the LFV processes from the point of view of the LRM.
Our goal is to consider the Higgs decay $H\to\mu\tau$ and
establish whether this decay is possible in the context of the
LRM. The organization of the paper goes as follows: section 2
contains a summary of the LRM. In sections 3 we fulfill our
calculations and analyze the results obtained. Section 4 includes
our conclusion.

\section{The left-right-symmetric model}

In the LRM quarks and leptons enter into the left- and
right-handed doublets
$$\left.\begin{array}{ll}
\displaystyle{Q_L^a({1\over2}, 0,
{1\over3})=\left(\matrix{u_L^a\cr d_L^a}\right)},\hspace{12mm}
\displaystyle{Q_R^a(0,{1\over2}, {1\over3})=\left(\matrix{u_R^a\cr
d_R^a}\right)},\\[4mm]
\displaystyle{\Psi_L^a({1\over2}, 0, -1)=\left(\matrix{\nu_{a
L}\cr l_{\alpha L}}\right)},\qquad
\displaystyle{\Psi_R^a(0,{1\over2}, -1)=\left(\matrix{N_{a R}\cr
l_{a R}}\right)},\end{array}\right\}\eqno(5)$$ where $a=1,2,3$, in
brackets the values of $S^W_L, S^W_R$ and $B-L$ are given, $S^W_L$
($S^W_R$) is the weak left (right) isospin while $B$ and $L$ are
the baryon and lepton numbers. Note that introducing the heavy
neutrinos $N_{aR}$ leads to the existence of the see-saw relation
which, in its turn, gives explanation of the $\nu_l$-neutrino mass
smallness. The Higgs sector structure of the LRM determines the
neutrino nature. The mandatory element of the Higgs sector is the
bi-doublet $\Phi(1/2,1/2,0)$
$$\Phi=\left(\matrix{\Phi^0_1 & \Phi^+_2\cr
\Phi^-_1 &\Phi^0_2\cr}\right).\eqno(6)$$ Its nonequal vacuum
expectation values (VEV's) of the electrically neutral components
bring into existence the masses of quarks and leptons. For the
neutrino to be a Majorana particle, the Higgs sector must include
two triplets $\Delta_L(1,0,2)$, $\Delta_R(0,1,2)$ \cite{RN81}
$$({\mbox{\boldmath $\tau$}}\cdot{\mbox{\boldmath $\Delta$}}_L)=
\left(\matrix{\delta_L^+/\sqrt{2} & \delta_L^{++}\cr \delta_L^0 &
-\delta_L^+/\sqrt{2}\cr}\right),\qquad ({\mbox{\boldmath
$\tau$}}\cdot{\mbox{\boldmath
$\Delta$}}_R)=\left(\matrix{\delta_R^+/\sqrt{2} & \delta_R^{++}\cr
\delta_R^0 & -\delta_R^+/\sqrt{2}\cr}\right).\eqno(7)$$ If the
Higgs sector consists of two doublets $\chi_L(1/2,0,1)$,
$\chi_R(0,1/2,1)$ and one bidoublet $\Phi(1/2,1/2,0)$
\cite{RMS77}, then the neutrino represents a Dirac particle. In
what follows we shall consider the LRM version with Majorana
neutrinos.

The masses of fermions and their interactions with the gauge boson
are controlled by the Yukawa Lagrangian. Its expression for the
lepton sector is as follows
$${\cal L}_Y=-\sum_{a,b}\{h_{ab}\overline{\Psi}_{aL}\Phi\Psi_{bR}+
h^{\prime}_{ab}\overline{\Psi}_{aL}\tilde{\Phi}\Psi_{b,R}+ $$
$$+if_{ab}[\Psi^T_{aL}C\tau_2({\mbox{\boldmath $\tau$}}\cdot
{\mbox{\boldmath $\Delta$}}_L)\Psi_{bL}+ (L\rightarrow
R)]+\mbox{h.c.}\},\eqno(8)$$ where $C$ is a charge conjugation
matrix, $\tilde{\Phi}= \tau_2\Phi^*\tau_2$, $a,b=e,\mu,\tau,$
$h_{ab}, h^{\prime}_{ab}$ and $f_{ab}=f_{ba}$ are bidoublet and
triplet Yukawa couplings (YC's), respectively.

The spontaneous symmetry breaking (SSB) according to the chain
$$SU(2)_L\times SU(2)_R\times U(1)_{B-L}\rightarrow SU(2)_L\times U(1)_Y
\rightarrow U(1)_Q$$ is realized for the following choice of the
vacuum expectation values (VEV's):
$$<\delta^0_{L,R}>={v_{L,R}\over\sqrt{2}},\qquad <\Phi^0_1>=k_1,\qquad
<\Phi^0_2>=k_2.\eqno(9)$$ To achieve agreement with experimental
data, it is necessary to ensure fulfillment of the conditions
$$v_L<<\mbox{max}(k_1,k_2)<<v_R.\eqno(10)$$

The Higgs potential $V_H$ is the essential element of the theory
because it defines the physical states basis of Higgs bosons,
Higgs masses, and interactions between Higgses. We shall use the
most general shape of $V_H$ that was proposed in Ref.
\cite{NGD91}. After the SSB we have 14 physical Higgs bosons. They
are: four doubly-charged scalars $\Delta^{(\pm)}_{1,2}$, four
singly-charged scalars $\tilde{\delta}^{(\pm)}$ and $h^{(\pm)}$,
four neutral scalars $S_{1,2,3,4}$ ($S_1$ boson is an analog of
the SM Higgs boson), and two neutral pseudoscalars $P_{1,2}$.

We now direct our attention to the sector of the neutral scalar
Higgses. If one does not impose any conditions on the constants
entering the Higgs potential $V_H$, then we have four scalars
$$\left.\begin{array}{ll}
S_1=(\Phi_-^{0r}\cos\theta_0+\Phi_+^{0r}\sin\theta_0)\cos\alpha
-\delta_R^{0r}\sin\alpha, \
S_2=-\Phi_-^{0r}\sin\theta_0+\Phi_+^{0r}\cos\theta_0,\\[2mm]
\hspace{15mm}S_3=(\Phi_-^{0r}\cos\theta_0+\Phi_+^{0r}\sin\theta_0)\sin\alpha
+\delta_R^{0r}\cos\alpha, \qquad
S_4=\delta^{0r}_L,\end{array}\right\}\eqno(11)$$ where
$$\Phi_-^{0r}={k_1\Phi_1^{0r}+k_2\Phi_2^{0r}\over k_+},\qquad
\Phi_+^{0r}={k_1\Phi_2^{0r}-k_2\Phi_1^{0r}\over k_+},$$
$k_{\pm}=\sqrt{k_1^2\pm k_2^2}$ and the superscript $r$ means the
real part of the corresponding quantity. The mixing angle
$\theta_0$ is defined by the expression \cite{Bom200}
$$\tan2\theta_0={{4k_1k_2k_-^2[2(2\lambda_2+\lambda_3)k_1k_2+\lambda_4k_+^2]}
\over{k_1k_2[(4\lambda_2+2\lambda_3)(k_-^4-4k_1^2k_2^2)-k_+^2
(2\lambda_1k_+^2+8\lambda_4k_1k_2)]-\alpha_2v_R^2k_+^4}}\eqno(12)$$
and, as a result, appears to be very small. In what follows we
shall set it equal to zero. As far as the mixing angle $\alpha$ is
concerned, it could be very sizeable. The theory predict that at
$v_L=k_2=0$ the expression for the mixing angle $\alpha$ is as
follows \cite{JG96}
$$\tan2\alpha={\alpha_Hk_1v_R\over\rho_Hv_R^2-\lambda_Hk_1^2},\eqno(13)$$
where $\lambda_H, \rho_H$ and $\alpha_H$ are linear combinations
of the constants entering the Higgs potential. Recent
investigations \cite{FA15,SIG16} allow for $\sin\alpha<0.44$ at
$2\sigma$ CL, practically independently of the $S_3$ mass. Then
the Lagrangian of interaction between the $S_1$ boson and leptons
will look like
$${\cal{L}}_l=-{1\over\sqrt{2}k_+}\Big\{\sum_am_a\overline{l}_{aR}l_{aL}
S_1\cos\alpha+\sum_{a,b}\overline{N}_{aR}\nu_{bL}[h_{ab}
k_1+h^{\prime}_{ab}k_2]S_1\cos\alpha\Big\}+\mbox{h.c.}.\eqno(14)$$

It is convenient to express the coupling constants of the $S_1$
boson with the neutrinos in terms of neutrino oscillation
parameters \cite{Bom200,Bom97}. In the two flavor approximation
the neutrino mass matrix in the basis
$\Psi^T=\left(\nu_{aL}^T,N_{aR}^T,\nu_{bL}^T,N_{bR}^T\right)$ will
look like
$${\cal M}=\left(\matrix{f_{aa}v_L &
m^a_D & f_{ab}v_L & M_D\cr m_D^a & f_{aa}v_R & M^{\prime}_D &
f_{ab}v_R\cr f_{ab}v_L & M^{\prime}_D & f_{bb}v_L & m^b_D \cr M_D
& f_{ab}v_R & m_D^b & f_{bb}v_R\cr}\right).\eqno(15)$$ where
$$m_D^a=h_{aa}k_1+h^{\prime}_{aa}k_2,\eqno(16)$$
$$M_D=h_{ab}k_1+h^{\prime}_{ab}k_2,\qquad
M^{\prime}_D=h_{ba}k_1+h^{\prime}_{ba}k_2.\eqno(17)$$ The
transition to the eigenstate neutrino mass basis $m_i$
($i=1,2,3,4$) is carried out by the matrix
$$U=\left(\matrix{c_{\varphi_a}c_{\theta_{\nu}}
& s_{\varphi_a}c_{\theta_N} & c_{\varphi_a}s_{\theta_{\nu}}  &
s_{\varphi_a}s_{\theta_N} \cr -s_{\varphi_a}c_{\theta_{\nu}} &
c_{\varphi_a}c_{\theta_N} & -s_{\varphi_a}s_{\theta_{\nu}}  &
c_{\varphi_a}s_{\theta_N} \cr -c_{\varphi_b}s_{\theta_{\nu}}
&-s_{\varphi_b}s_{\theta_N} & c_{\varphi_b}c_{\theta_{\nu}} &
s_{\varphi_b}c_{\theta_N}\cr s_{\varphi_b}s_{\theta_{\nu}}
&-c_{\varphi_b}s_{\theta_N} & -s_{\varphi_b}c_{\theta_{\nu}} &
c_{\varphi_b}c_{\theta_N}\cr}\right),\eqno(18)$$ where $\varphi_a$
and $\varphi_b$ are the mixing angles inside $a$ and $b$
generations respectively, $\theta_{\nu} (\theta_N)$ is the mixing
angle between the light (heavy) neutrinos belonging to the $a$-
and $b$-generations, $c_{\varphi_a}=\cos\varphi_a, \
s_{\varphi_a}=\sin\varphi_a$ and so on. Using the eigenvalues
equation for the mass matrix we could obtain the relations which
connect the YC's with the masses and mixing angles of the
neutrinos
$$m_D^a=c_{\varphi_a}s_{\varphi_a}(-m_1c^2_{\theta_{\nu}}-m_3s^2_
{\theta_{\nu}}+ m_2c^2_{\theta_N}+m_4s^2_{\theta_N}),\eqno(19)$$
$$M_D=c_{\varphi_a}s_{\varphi_b}c_{\theta_{\nu}}s_{\theta_{\nu}}
(m_1-m_3)+s_{\varphi_a}c_{\varphi_b}c_{\theta_N}s_{\theta_N}(m_4-m_2),
\eqno(20)$$
$$f_{ab}v_R=s_{\varphi_a}s_{\varphi_b}c_{\theta_{\nu}}s_{\theta_{\nu}}
(m_3-m_1)+c_{\varphi_a}c_{\varphi_b}c_{\theta_N}s_{\theta_N}(m_4-m_2),
\eqno(21)$$
$$f_{aa}v_R=(s_{\varphi_a}c_{\theta_{\nu}})^2m_1+(c_{\varphi_a}c_{\theta_N})
^2m_2+(s_{\varphi_a}s_{\theta_{\nu}})^2m_3+(c_{\varphi_a}s_{\theta_N})^2m_4,
\eqno(22)$$
$$f_{bb}v_R=(s_{\varphi_b}s_{\theta_{\nu}})^2m_1+(c_{\varphi_b}
s_{\theta_b})^2m_2+(s_{\varphi_b}c_{\theta_{\nu}})^2m_3+
(c_{\varphi_b}c_{\theta_N})^2m_4,\eqno(23)$$
$$ m_D^b=m_D^a(\varphi_a\rightarrow \varphi_b,\theta_{\nu,N}
\rightarrow\theta_{\nu,N}+{\pi\over2}),\qquad
M_D^{\prime}=M_D(\varphi_a\leftrightarrow\varphi_b),\eqno(24)$$
The change $L\rightarrow R$ in the left-hand sides of Eqs.
(21)-(23) results in the replacement
$\varphi_{a,b}\rightarrow\varphi_{a,b}+{\pi\over2}$ in their
right-hand sides. From definition of $f_{aa}v_R$ and $f_{aa}v_L$
follows the exact formula for the heavy-light neutrino mixing
angle $\varphi_{a,b}$ \cite{Bom04}
$$\sin2\varphi_a=2{\sqrt{f^2_{aa}v_Rv_L-[f_{aa}(v_R+v_L)-
m_{\nu_1}c_{\theta_{\nu}}^2-m_{\nu_2}s_{\theta_{\nu}}^2]
(m_{\nu_1}c_{\theta_{\nu}}^2+m_{\nu_2}s_{\theta_{\nu}}^2)} \over
f_{aa}(v_R+v_L)-2(m_{\nu_1}c_{\theta_{\nu}}^2+
m_{\nu_2}s_{\theta_{\nu}}^2)},\eqno(25)$$
$$\sin2\varphi_b=\sin2\varphi_a\left(f_{aa}\rightarrow f_{bb},
\theta_{\nu}\rightarrow\theta_{\nu}+{\pi\over2}\right).\eqno(26)$$
It should be remarked that according the LRM the heavy-light
mixing angles belonging to different generations are practically
equal in value
$$\sin2\varphi_a\simeq\sin2\varphi_b\simeq2{\sqrt{v_Rv_L}\over
v_R+v_L}\equiv\sin2\varphi.\eqno(27)$$

In following calculations we also need the Lagrangians which
describe interaction of the charged gauge bosons both with the
$S_1$ Higgs boson
$$\sqrt{2}{\cal L}^n_W=g_L^2\Big\{k_+[W^*_{1\mu}(x)W_1^{\mu}(x)+
W^*_{2\mu}(x)W_2^{\mu}(x)]-{2k_1k_2\over k_+}[c_{2\xi}
(W^*_{2\mu}(x)W_1^{\mu}(x)+W^*_{1\mu}(x)W_2^{\mu}(x))+$$
$$+s_{2\xi} (W^*_{2\mu}(x)W_{2\mu}(x)-
W^*_{1\mu}(x)W_{1\mu}(x))]\Big\}S_1(x),\eqno(28)$$ and with
leptons
$${\cal{L}}_l^{CC}={g_L\over2\sqrt{2}}\sum_l\Big[\overline{l}(x)\gamma^{\mu}
(1-\gamma_5) \nu_{lL}(x)W_{L\mu}(x)+\overline{l}(x)\gamma^{\mu}
(1+\gamma_5)N_{lR}(x)W_{R\mu}(x)\Big],\eqno(29)$$ where
$$W_1=W_L\cos\xi+W_R\sin\xi,\qquad W_2=-W_L\sin\xi+W_R\cos\xi,$$
The theory predicts the following connection between the heavy
charged gauge boson mass $m_{W_2}$ ($m_{W_2}\simeq g_Lv_R$) and
the mixing angle $\xi$ \cite{RN81}
$$\tan2\xi\simeq{4g_Lg_Rk_1k_2\over
g_R^2(2v_R^2+k_+^2)-g_L^2(2v_L^2+k_+^2)}.\eqno(30)$$

In Ref. \cite{OMB14} investigation of Mikheyev-Smirnov-Wolfenstein
resonance with the solar and reactor neutrinos has be done. The
sector of heavy neutrino in two flavor approximation has been
considered. It was demonstrated that only three versions of the
heavy neutrino sector structure are possible: (i) the light-heavy
neutrino mixing angles $\varphi_a$ and $\varphi_b$ are arbitrary
but equal each other whereas the heavy neutrino masses are
quasi-degenerate (quasi-degenerate mass case --- QDM case); (ii)
the heavy neutrino masses are hierarchical ($m_{N_1}< m_{N_2}$)
while the angles $\varphi_a$ and $\varphi_b$ are equal to zero (no
mass degeneration case --- NMD case); (iii) $\varphi_a=\varphi_b$
and the heavy-heavy neutrino mixing is maximal, $\theta_N=\pi/4$,
and as a result the heavy neutrino masses are hierarchical
(maximal heavy-heavy mixing case
--- MHHM case). It is logical to assume that the same pattern
takes place in the three flavor approximation as well.

\section{Decay of the Higgs boson into $\mu\tau$ pair}

In this chapter we shall investigate the Higgs decay into the
channel
$$S_1
\to \mu^++\tau^-\eqno(31)$$ within the LRM. Thanks to the mixing
into the neutrino sector this decay could go in the third order of
the perturbation theory. The corresponding diagrams are pictured
in Fig.1.
\begin{figure}[!h]
\begin{center}
\includegraphics[width =0.7\textwidth]{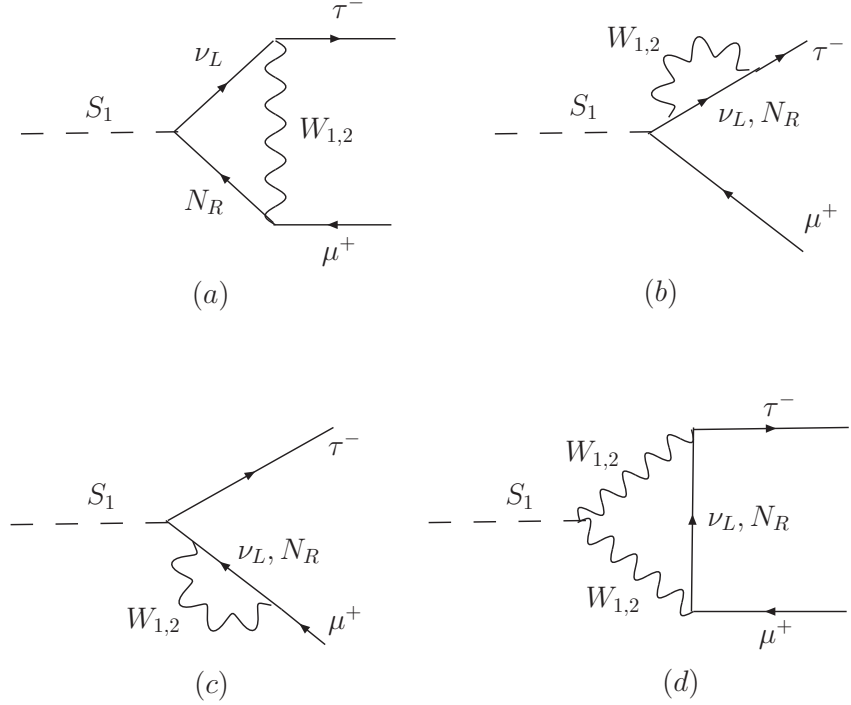}
\end{center}
\caption {The Feynman diagrams contributing to the decay
$S_1\to \mu^++\tau^-$.}
\end{figure}
For the sake od simplicity we shall consider the individual
contributions of each diagram to the total width of the decay
(31). Let us start with the kind of the diagrams one of them shown
in Fig.1a. There are eight diagrams depending on what neutrinos
are produced in the virtual state. For example, when in the
virtual state the $\nu_{\tau}\overline{N}_{\tau}$ pair comes into
being the corresponding matrix element take the form
$$M_1^{(a)}={g_L^2m_D^{\tau}\cos\alpha\sin2\theta_N\sin\xi\over
32k_+\sqrt{2}}\sqrt{{m_{\tau}m_{\mu}\over2m_{S_1}E_{\tau}E_{\mu}}}\
\overline{u}(p_1)\gamma_{\lambda}(1-\gamma_5)\Big\{\int_{\Omega}
{\hat{p}-\hat{k}+m_{\nu_i}\over(p-k)^2-m_{\nu_i}^2} \times$$
$$\times(1+\gamma_5)\Bigg[{\hat{k}+m_{N_2}\over k^2-m_{N_2}^2}
-{\hat{k}+m_{N_1}\over
k^2-m_{N_1}^2}\Bigg]\gamma_{\sigma}(1+\gamma_5)
{g^{\lambda\sigma}-(k-p_2)^{\lambda}(k-p_2)^{\sigma}/m_{W_1}^2\over
(k-p_2)^2-m_{W_1}^2} d^4k\Big\}v(p_2),\eqno(32)$$ where $m_{N_j}$
$(j=1,2)$ is the mass of the heavy neutrino, $p_1$ and $p_2$ are
momentum of $\tau$-lepton and $\mu$-meson, respectively. Taking
into account Eqs. (19), (20) and (24) we find that the matrix
element corresponding to all eight diagrams is given by the
expression
$$M^{(a)}=\sum_{i=1}^8M_i^{(a)}={g_L^2\cos\alpha\sin2\varphi\sin2
\theta_N\sin\xi\over
16k_+\sqrt{2}}\sqrt{{m_{\tau}m_{\mu}\over2m_{S_1}E_{\tau}E_{\mu}}}\
\overline{u}(p_1)\gamma_{\lambda}(1-\gamma_5)\Big\{\int_{\Omega}
{\hat{p}-\hat{k}+m_{\nu_i}\over(p-k)^2-m_{\nu_i}^2} \times$$
$$\times(1+\gamma_5)\Bigg[{m_{N_2}(\hat{k}+m_{N_2})\over k^2-m_{N_2}^2}
-{m_{N_1}(\hat{k}+m_{N_1})\over
k^2-m_{N_1}^2}\Bigg]\gamma_{\sigma}(1+\gamma_5)
{g^{\lambda\sigma}-(k-p_2)^{\lambda}(k-p_2)^{\sigma}/m_{W_1}^2\over
(k-p_2)^2-m_{W_1}^2} d^4k\Big\}v(p_2).\eqno(33)$$

Substituting (33) into the partial decay width
$$d\Gamma=(2\pi)^4\delta^{(4)}(p-p_1-p_2)|M^{(a)}|^2{d^3p_1d^3p_2
\over(2\pi)^8},$$ integrating the obtained expression over $p_1$,
$p_2$ and using the procedure of dimensional regularization, we
get
$$\Gamma(S_1\to\overline{\nu}_L^*N_R^*W_1^*\to\mu^+\tau^-)=
{\pi^3(g_L^2\cos\alpha\sin2\varphi\sin2\theta_N\sin\xi)^2
\over16m_{S_1}^3}\Big\{4m_{\tau}m_{\mu}(\Delta L)(\Delta R)+$$
$$+(m_{S_1}^2-m_{\tau}^2-m_{\mu}^2)[(\Delta L)^2+(\Delta
R)^2]\Big\}\sqrt{(m_{S_1}^2-m_{\mu}^2-m_{\tau}^2)^2-4m_{\mu}^2m_{\tau}^2}
,\eqno(34)$$ where
$$\Delta L=L(m_{N_2})-L(m_{N_1}),\qquad
L(m_{N_j})={m_{N_j}\over
k_+}\Big[L_W^1(m_{N_j})+L_W^2(m_{N_j})+L_W^3(m_{N_j})\Big],$$
$$\Delta R=R(m_{N_2})-R(m_{N_1}),\qquad
R(m_{N_j})={m_{N_j}\over
k_+}\Big[R_g(m_{N_j})+R^1_W(m_{N_j})+R^2_W(m_{N_j})+$$
$$+R^3_W(m_{N_j})+ R^4_W(m_{N_j})\Big],$$
$$R_g(m_{N_j})=2\int_0^1xdx\int_0^1\Big[
{(pp_x)-p_x^2\over l_{xy}^j-p_x^2} -2\ln\Bigg|{l_{xy}^j\over
l_{xy}^j-p_x^2}\Bigg|\Big]dy,\eqno(35)$$
$$L_W^1(m_{N_j})={2m_{\mu}m_{\tau}\over
m_W^2}\int_0^1xdx\int_0^1{(m_{S_1}^2-m_{\tau}^2)(x-xy)-2(p_2p_x)+m_{\mu}^2x\over
l_{xy}^j-p_x^2}dy,\eqno(36)$$
$$R_W^1(m_{N_j})=-{2m^2_{\mu}\over
m_W^2}\int_0^1xdx\int_0^1{(m_{S_1}^2-m_{\tau}^2)x+m_{\tau}^2(x-xy)\over
l_{xy}^j-p_x^2}dy,\eqno(37)$$
$$L_W^2(m_{N_j})=-{2m_{\mu}m_{\tau}\over
m_W^2}\int_0^1xdx\int_0^1\Bigg[-3\ln\Bigg|{l_{xy}^j\over
l_{xy}^j-p_x^2}\Bigg|+{(pp_x)(x-xy)-2p_x^2\over
l_{xy}^j-p_x^2}\Bigg]dy,\eqno(38)$$
$$R_W^2(m_{N_j})={2\over
m_W^2}\int_0^1xdx\int_0^1\Bigg[\ln\Bigg|{l_{xy}^j\over
l_{xy}^j-p_x^2}\Bigg|(2m_{S_1}^2-2m_{\tau}^2+m_{\mu}^2)+{(pp_x)xm_{\mu}^2+
(m_{S_1}^2- m_{\tau}^2)p_x^2\over
l_{xy}^j-p_x^2}\Bigg]dy,\eqno(39)$$
$$L_W^3(m_{N_j})=-{m_{\mu}m_{\tau}\over
m_W^2}\int_0^1xdx\int_0^1\Bigg\{6xy\ln\Bigg|{l_{xy}^j\over
l_{xy}^j-p_x^2}\Bigg|+{(2xy-4x)p_x^2\over
l_{xy}^j-p_x^2}\Bigg\}dy,\eqno(40)$$
$$R_W^3(m_{N_j})=-{1\over
m_W^2}\int_0^1xdx\int_0^1\Bigg\{\ln\Bigg|{l_{xy}^j\over
l_{xy}^j-p_x^2}\Bigg|\Bigg[12(pp_x)+6m_{\mu}^2x-6m_{\tau}^2(x-xy)\Bigg]+
$$ $$+{2p_x^2\over
l_{xy}^j-p_x^2}\Bigg[2(pp_x)+m_{\mu}^2x-m_{\tau}^2(x-xy)\Bigg]\Bigg\}dy,
\eqno(41)$$
$$R_W^4(m_{N_j})={1\over
m_W^2}\int_0^1xdx\int_0^1\Bigg\{\ln\Bigg|{l_{xy}^j\over
l_{xy}^j-p_x^2}\Bigg|(24p^2_x-12l_{xy}^j)
+p_x^2\Bigg[12+{2p_x^2\over
l_{xy}^j-p_x^2}\Bigg]\Bigg\}dy,\eqno(42)$$
$$l_{xy}^j=yx(m_{\mu}^2-m_{W_1}^2-m_{S_1}^2)+x(m_{S_1}^2+m_{N_j}^2)-m_{N_j}^2,$$
$$p_x^2=m_{\tau}^2x^2y^2+m_{S_1}^2x^2-(m_{S_1}^2+m_{\tau}^2-m_{\mu}^2)x^2y,\qquad
(pp_x)=m_{S_1}^2x-{1\over2}(m_{S_1}^2-m_{\mu}^2+m_{\tau}^2)xy,$$
$$(p_2p_x)=m_{\mu}^2x+{1\over2}(m_{S_1}^2-m_{\mu}^2-m_{\tau}^2)(x-xy),$$
In the expression (33) we have neglected mixing in the light
neutrino sector because of current experiments leads to the
results \cite{CP16}
$$\Delta(m_{21})^2=\mbox{few}\times10^{-5}\
\mbox{eV}^2,\qquad \Delta(m_{31})^2=\mbox{few}\times10^{-3}\
\mbox{eV}^2,\qquad\Delta(m_{32})^2=\mbox{few}\times10^{-3}\
\mbox{eV}^2.\eqno(43)$$

Now we proceed to the diagrams of Fig.1b-1d. Calculations show
that amongst them the greatest contributions are come from the
following two diagrams pictured on Fig.1d. The first diagram
contains the $W^-_2W_2^+N_R$ particles in the virtual states. Its
existence is caused by the heavy-heavy neutrino mixing (HHNM) and,
as a result, contribution from this diagram turns into zero when
$\theta_N=0$. The second diagram holds the $W_1^-W_1^+\nu_L$
particles in the virtual states and it leads to nonzero
contribution in only case when both the HHNM and the heavy-light
neutrino mixing are in existence. It is convenient to consider
contributions of these diagrams to the decay width separately. In
the case of the HHNM we obtain
$$\Gamma(S_1\to W_2^{+*}W_2^{-*}N_R^*\to\mu^+\tau^-)=
{\pi^3(g_L^4k_+\sin2\theta_N)^2\over128m_{S_1}^3}\Big\{4m_{\tau}
m_{\mu}(\Delta L^{\prime})(\Delta
R^{\prime})+(m_{S_1}^2-m_{\tau}^2-$$ $$-m_{\mu}^2)[(\Delta
L^{\prime})^2+(\Delta
R^{\prime})^2]\Big\}\sqrt{(m_{S_1}^2-m_{\mu}^2-m_{\tau}^2)^2-
4m_{\mu}^2m_{\tau}^2} ,\eqno(44)$$ where the expressions for
$\Delta L^{\prime}$ and $\Delta R^{\prime}$ are given in Appendix.

The expression for $\Gamma(S_1\to
W_1^{+*}W_1^{-*}\nu_L^*\to\mu^+\tau^-)$ follows from (44) under
replacement
$$m_{W_2}\to
m_{W_1},\qquad(\sin2\theta_N)^2\to(\sin2\theta_N\sin^2\varphi)^2.\eqno(45)$$

In order to compare the obtained expressions it is necessary to
have information concerning the values of such parameters as
$v_R$, $\xi$, $v_L$ and $\varphi$. Let us start with the $v_R$ and
$\xi$. The lower bound obtained by the ATLAS Collaboration on
$m_{W_2}$ from dijet searches at $\sqrt{s}=13$ TeV is \cite{ATC17}
$$m_{W_2}\geq3.7\ \mbox{TeV}\qquad\mbox{at}\ 95\%
\mbox{C.L.}\qquad\mbox{with}\qquad L=37\
\mbox{fb}^{-1},\eqno(46)$$ to give $v_R\simeq5.7$ TeV. Since
current experimental limits on the mixing angle $\xi$ fall in the
broad range between 0.12 and 0.0006 (see, for review \cite{CP16}),
then for definition of $\xi$ one needs to use the relation (30)
which is predicted by the LRM. Using $v_R=5.7$ TeV we get
$\xi\simeq5\times10^{-4}$. In what follows we shall use this very
value for the mixing angle $\xi$.

As far as the value of the heavy-light neutrino mixing angle
$\varphi$ is concerned, there are a lot of papers devoted to
determination of experimental bounds on it (see, for example
\cite{PSB13} and references therein). One way to find such bounds
is connected with searches for the neutrinoless double beta decay
($0\nu\beta\beta$) and disentangle the heavy neutrino effect. In
Ref. \cite{SD16} considering the case of $^{76}\mbox{Ge}$, the
following expression was obtained
$$\Big|\sum_i{U_{ei}^2\over m_{N_i}}\Big|<{7.8\times10^{-8}\over
m_p}\Bigg[{104\over
{\cal{M}}_{0\nu}(\mbox{Ge})}\Bigg]\times\Bigg[{3\times10^{25}\
\mbox{yr}\over \tau^{0\nu}_{1/2}}\Bigg]^{1/2},\eqno(47)$$ where
${\cal{M}}_{0\nu}(\mbox{Ge})$ is is the nuclear matrix element,
$m_p$ is the proton mass and $\tau^{0\nu}_{1/2}$ is the half-life
for $0\nu\beta\beta$. However, there is the point of view that the
$0\nu\beta\beta$ does not give the reliable answer on the value of
the heavy-light mixing. Of course, the main uncertainties are
connected with the determination of nuclear matrix element. In its
calculation one should assume the definite values both for the
axial coupling constants of the nucleon $g_A$ and for the phase
space factor. For example, when
$g_A=g_{\mbox{\footnotesize{nucleon}}}=1.269$\qquad and
$g_A=g_{\mbox{\footnotesize{phen.}}}=g_{\mbox
{\footnotesize{nucleon}}}\times A^{-0.18}$ ($A$ is the atomic
number) the ${\cal{M}}_{0\nu}(\mbox{Ge})$ takes the values
$104\pm29$ and $22\pm6$, respectively. Note, the
$g_A=g_{\mbox{\footnotesize{phen.}}}$ parametrization as a
function of $A$ comes directly from the comparison between the
theoretical half-life for $2\nu\beta\beta$ and its observation in
different nuclei \cite{JB13}. Using
$\tau^{0\nu}_{1/2}(^{76}\mbox{Ge})=1.9\times10^{25}$ yr and
setting $m_N=100$ GeV, with the help of Eq. (47) we may get
$$(\sin\varphi)_{\mbox{\footnotesize{max}}}\simeq\left\{\begin{array}
{ll}3.2\times10^{-3}\qquad\mbox{when} \qquad
g_A=g_{\mbox{\footnotesize{nucleon}}} ,\\
7\times10^{-3}\qquad\hspace{4mm}\mbox{when}\qquad
g_A=g_{\mbox{\footnotesize{phen.}}} .\end{array}\right.$$

The other way is to directly look for the presence of the
heavy-light neutrino mixing, which can manifest in several ways,
for example, (i) via departures from unitarity of the neutrino
mixing matrix, which could be investigated in neutrino oscillation
experiments as well as in lepton flavor violation searches, and
(ii) via their signatures in collider experiments. To take an
illustration, in Ref. \cite{CC13} the final states with same-sign
dileptons plus two jets without missing energy
($l^{\pm}l^{\pm}jj$), arising from $pp$ collisions were
considered. This signal depends crucially on the heavy-light
neutrino mixing. Analysis of the channel
$$p+p\to N^*_ll^{\pm}\to l^{\pm}+l^{\pm}+2j\eqno(48)$$
led to the upper limit on $\sin\varphi$ equal to
$3.32\times10^{-2}$ for $m_{W_R}=4$ TeV and $m_{N_l}=100$ GeV. On
the other hand to evaluate $\varphi$ we could use the relation
(27) as well. The precision measurements of the electroweak $\rho$
parameter \cite{TGR82}
$$\rho={m_{Z_1}^2\cos^2\theta_W\over
m_{W_1}^2}={1+4x\over1+2x}\eqno(49)$$ ($x=(v_L/k_+)^2$) set an
upper bound on the VEV of $v_L\leq3$ GeV. Taking into account this
value we obtain
$$(\sin2\varphi)_{\mbox{\footnotesize{max}}}\simeq4.6\times10^{-2}.\eqno(50)$$

Setting
$$\left.\begin{array}{ll}
\theta_N={\pi\over4},\qquad m_{N_1}=140\ \mbox{GeV},\qquad
m_{N_2}=250\ \mbox{GeV},\\  [2mm] \sin\alpha=0.44,\qquad
\sin\xi=5\times10^{-4},\qquad \sin\varphi=2.3\times10^{-2},
\end{array}\right\}\eqno(51)$$
we get
$${\Gamma(S_1\to
\nu_L^*N_R^*W_1^*\to\mu^+\tau^-)\over \Gamma(S_1\to
W_1^*W_1^*\nu_L^*\to\mu^+\tau^-)}\simeq10^5,\qquad {\Gamma(S_1\to
\nu_L^*N_R^*W_1^*\to\mu^+\tau^-)\over \Gamma(S_1\to
W_2^*W_2^*N_R^*\to\mu^+\tau^-)}\simeq10^4.\eqno(52)$$ So, the main
contribution to the decay $S_1\to\mu^++\tau^-$ comes from the
diagram of Fig.1a.

In order to obtain the width of the decay
$$S_1\to\mu^-+\tau^+\eqno(53)$$
one should make in Eqs. (34) the following replacement
$$m_{\tau}\leftrightarrow m_{\mu}.$$
Now we shall find out whether could the obtained expressions for
$\mbox{BR}(S_1\to\mu^+\tau^-)+\mbox{BR}(S_1\to\mu^-\tau^+)$
reproduce the experimental bound on the branching ratio of the
decay $S_1\to\mu\tau$? First and foremost we note that the width
of this decay does not equal to zero only provided the heavy
neutrino masses are hierarchical while the heavy-heavy and
heavy-light neutrino mixing angles do not equal to zero. Using
(51) we get
$$\mbox{BR}(S_1\to\tau^-\mu^+)\simeq\Bigg\{\begin{array}{ll}
0.24\times10^{-4},\qquad\mbox{when}\qquad\sin\varphi=2.3
\times10^{-2},\\
[2mm]
0.45\times10^{-6},\qquad\mbox{when}\qquad\sin\varphi=3.2\times10^{-3}
.\end{array}.\eqno(54)$$ So, we see that at most the obtained
expression is two orders of magnitude less than the current
experimental upper bound being equal to $0.25\times10^{-2}$.

\section{Conclusion}

Within the left-right symmetric model (LRM) the decays of the
neutral Higgs boson $S_1$
$$S_1\to\mu^++\tau^-,\qquad S_1\to\mu^-+\tau^+\eqno(55)$$
where $S_1$ is an analog of the standard model (SM) Higgs boson,
have been considered. These decays go with the lepton flavor
violation (LFV) and, as result, are forbidden in the SM.

We have found the widths of the decays (55) in the third order of
the perturbation theory. The width of this decay does not equal to
zero only provided the heavy neutrino masses are hierarchical. It
was shown that the main contribution to the decay width is caused
by the diagram with the light and heavy neutrinos in the virtual
state. Therefore, investigation of these decays could give
information about the neutrino sector structure of the model under
study.

The obtained decay widths critically depend on the angle $\xi$
which defines the mixing in the charged gauge boson sector and the
heavy-light neutrino mixing angle $\varphi$. Within the LRM there
exist the formulae connecting the values of these angles with the
VEV's $v_L$ and $v_R$. Using the results of the current
experiments, on looking for the additional charged gauge boson
$W_2$ and on measuring the electroweak $\rho$ parameter, gives
$$\sin\xi\leq5\times10^{-4},\qquad\sin\varphi\leq2.3\times10^{-2}.
\eqno(56)$$ However, even using the upper bounds on $\sin\xi$ and
$\sin\varphi$ one does not manage to get for the branching ratio
$\mbox{BR}(S_1\to\tau\mu)$ the value being equal to upper
experimental bound $0.25\times10^{-2}$. The theoretical expression
for the branching ratio of the decay $S_1\to\tau\mu$ proves to be
on two orders of magnitude less than the upper experimental bound.
On the other hand, it should be remembered that in our case
$\mbox{BR}(S_1\to\tau\mu)_{exp}$ is nothing more than the
experiment precision limit, rather than the measured value of the
branching ratio. Therefore, the experimental programs with higher
precision than at present are required to get more detail
information about the decay $S_1\to\tau\mu$.

At future hadronic and leptonic colliders the more high statistics
of Higgs boson events will be achieved. For example, the future
LHC runs with $\sqrt{s}=14$ TeV and total integrated luminosity of
first 300 $\mbox{fb}^{-1}$ and later 3000 $\mbox{fb}^{-1}$ expect
the production of about 25 and 250 millions of Higgs boson events,
respectively, to be compared with 1 million Higgs boson events
that the LHC produced after the first runs \cite{ATLAS1, CMS1}.
These large numbers provide an upgrading of sensitivities to
$\mbox{BR}(S_1\to l_k\overline{l}_m)_{exp}$ of at least two orders
of magnitude with respect to the present sensitivity. In much the
same way, at the planned lepton colliders, similar to the
international linear collider with $\sqrt{s}=1$ TeV and
$\sqrt{s}=2.5$ TeV \cite{BBF13}, and the future electron-positron
circular collider, formerly known as TLEP, with $\sqrt{s}=350$ GeV
and 10 $\mbox{ab}^{-1}$ \cite{TLEP}, the expectations are of about
1 and 2 million Higgs boson events, respectively, with much lower
backgrounds owing to the cleaner environment, which will also
allow for a large improvement in LFV Higgs boson decay searches
regarding to the current sensitivities.

\section*{Acknowledgments}

This work is partially supported by the grant of Belorussian
Ministry of Education No 20170217.

\section*{Appendix}
The terms appearing in the width of the decay
$$S_1\to W_2^{+*}W_2^{-*}N_R^*\to\mu^+\tau^-$$
are as follows:
$$\Delta L^{\prime}=L^{\prime}(m_{N_2})-L^{\prime}(m_{N_1}),\qquad
L^{\prime}(m_{N_j})=L_g^{\prime}(m_{N_j})+\sum_{i=1}^7L_W^{\prime
i}(m_{N_j}),$$
$$\Delta R^{\prime}=R^{\prime}(m_{N_2})-R^{\prime}(m_{N_1}),\qquad
R^{\prime}(m_{N_j})=R_g^{\prime}(m_{N_j})+\sum_{i=1}^7R_W^{\prime
i}(m_{N_j}),$$
$$L^{\prime}_g(m_{N_j})=2m_{\mu}\int_0^1x(x-1)dx\int_0^1
{dy\over \beta_{xy}^j-q_x^2},\qquad
R^{\prime}_g(m_{N_j})=-2m_{\tau}\int_0^1x^2dx\int_0^1 {ydy\over
\beta_{xy}^j-q_x^2},\eqno(A.1)$$
$$R^{\prime1}_W(m_{N_j})=-{m_{\tau}\over m_W^2}\int_0^1x^2dx\int_0^1ydy\Big\{6\ln\Big|
{l^j_{xy}\over l^j_{xy}-p_x^2}\Big|+2[p_x^2-2(p_xp_2)]{1\over
l^j_{xy}-p_x^2}\Big\},\eqno(A.2)$$
$$L^{\prime1}_W(m_{N_j})={2m_{\mu}\over m_W^2}\int_0^1xdx\int_0^1dy\Big\{(3x+1)\ln\Big|
{l^j_{xy}\over l^j_{xy}-p_x^2}\Big|+[p_x^2(x+1)-2(p_xp_2)x]{1\over
l^j_{xy}-p_x^2}\Big\},\eqno(A.3)$$
$$R^{\prime2}_W(m_{N_j})={m_{\tau}\over
m_W^2}\int_0^1xdx\int_0^1dy
\Big\{m_{S_1}^2xy+[(m_{\mu}^2-m_{S_1}^2-m_{\tau}^2)xy+
(m_{\tau}^2-m_{S_1}^2-m_{\mu}^2)(x-1)]\Big\}{1\over
l^j_{xy}-p_x^2},\eqno(A.4)$$
$$L^{\prime2}_W(m_{N_j})=-{m_{\mu}\over
m_W^2}\int_0^1xdx\int_0^1dy\Big\{ m_{S_1}^2(x-1)+
[(m_{\mu}^2-m_{S_1}^2-m_{\tau}^2)xy+
(m_{\tau}^2-m_{S_1}^2-m_{\mu}^2)(x-1)\Big\}{1\over
l^j_{xy}-p_x^2},\eqno(A.5)$$
$$R^{\prime3}_W(m_{N_j})={m_{\tau}\over
m_W^2}\int_0^1xdx\int_0^1dy\Big[4\ln\Big|{l_{xy}^j\over
l_{xy}^j-p_x^2}\Big|+{2p_x^2-2(p_xp_2)-2(pp_2)xy\over
l_{xy}^j-p_x^2}\Big],\eqno(A.6)$$
$$L^{\prime3}_W(m_{N_j})=-{m_{\mu}\over
m_W^2}\int_0^1xdx\int_0^1dy\Big[4\ln\Big|{l_{xy}^j\over
l_{xy}^j-p_x^2}\Big|+{2p_x^2-2(p_xp_2)-2(pp_2)x+2(p_xp)\over
l_{xy}^j-p_x^2}\Big],\eqno(A.7)$$
$$R^{\prime4}_W(m_{N_j})={m_{\tau}\over4m_W^4}\int_0^1x^2dx\int_0^1ydy\big\{
[80p_x^2-48l_{xy}^j-32(p_xp_2)]\ln\Big|{l_{xy}^j\over
l_{xy}^j-p_x^2}\Big|+$$ $$+{p_x^2\over
l_{xy}^j-p_x^2}[4p_x^2-8(p_xp_2)] \Big\},\eqno(A.8)$$
$$L^{\prime4}_W(m_{N_j})=-{m_{\mu}\over4m_W^4}\int_0^1xdx\int_0^1dy\Big\{
[(80p_x^2-48l_{xy}^j)x+32p_x^2-12l_{xy}^j-32(p_xp_2)x]
\ln\Big|{l_{xy}^j\over l_{xy}^j-p_x^2}\Big|+$$
$$+12p_x^2+{4p_x^2\over
l_{xy}^j-p_x^2}[p_x^2(x+1)-2(p_xp_2)x]\Big\},\eqno(A.9)$$
$$R^{\prime5}_W(m_{N_j})={m_{\tau}\over2m_W^4}\int_0^1xdx\int_0^1dy\big\{
[12l_{xy}^j-24p_x^2+6(m_{S_1}^2-m_{\tau}^2)xy+6m_{\mu}^2x]
\ln\Big|{l_{xy}^j\over l_{xy}^j-p_x^2}\Big|+$$
$$+{2p_x^2\over
l_{xy}^j-p_x^2}[(m_{S_1}^2-m_{\tau}^2)xy+m_{\mu}^2x-p_x^2]
-12p_x^2\Big\},\eqno(A.10)$$
$$L^{\prime5}_W(m_{N_j})={m_{\mu}\over2m_W^4}\int_0^1xdx\int_0^1dy\Big\{
[24p_x^2-12l_{xy}^j+6m_{\tau}^2xy-6m_{\mu}^2x]
\ln\Big|{l_{xy}^j\over l_{xy}^j-p_x^2}\Big|+$$
$$+12p_x^2+{2p_x^2\over
l_{xy}^j-p_x^2}[p_x^2+m_{\tau}^2xy-m_{\mu}^2x]\Big\}.\eqno(A.11)$$
$$R^{\prime6}_W(m_{N_j})={m_{\tau}\over2m_W^4}\int_0^1xdx\int_0^1dy\Big\{
[3l_{xy}^j-4p_x^2-8(p_xp)xy+2(p_xp_2)+2(pp_2)xy]
\ln\Big|{l_{xy}^j\over l_{xy}^j-p_x^2}\Big|+$$
$$+{2(p_xp)\over
l_{xy}^j-p_x^2}[2(p_xp_2)-p_x^2]xy-3p_x^2\Big\},\eqno(A.12)$$
$$L^{\prime6}_W(m_{N_j})={m_{\mu}\over2m_W^4}\int_0^1xdx\int_0^1dy\Big\{
[4p_x^2-3l_{xy}^j+8(p_xp)x-2(p_xp_2)-2(pp_2)x+4(p_xp)]
\ln\Big|{l_{xy}^j\over l_{xy}^j-p_x^2}\Big|+$$
$$+3p_x^2+{2(p_xp)\over
l_{xy}^j-p_x^2}[p_x^2x+p_x^2-2(p_xp_2)x]\Big\},\eqno(A.13)$$
$$R^{\prime7}_W(m_{N_j})={m_{\tau}\over2m_W^4}\int_0^1xdx\int_0^1dy\Big\{
[6(p_xp)-m_{S_1}^2-m_{\mu}^2+m_{\tau}^2]\ln\Big|{l_{xy}^j\over
l_{xy}^j-p_x^2}\Big|-$$
$$-{2(p_xp)\over
l_{xy}^j-p_x^2}[(m_{S_1}^2-m_{\tau}^2)xy+m_{\mu}^2x]\Big\},\eqno(A.14)$$
$$L^{\prime7}_W(m_{N_j})=-{m_{\mu}\over2m_W^4}\int_0^1xdx\int_0^1dy\Big\{
[6(p_xp)+m_{\tau}^2-m_{\mu}^2]\ln\Big|{l_{xy}^j\over
l_{xy}^j-p_x^2}\Big|+$$
$$+{2(p_xp)\over
l_{xy}^j-p_x^2}[m_{\tau}^2xy-m_{\mu}^2x]\Big\},\eqno(A.15)$$
$$\beta_{xy}^j=yx(m_{S_1}^2-m_{W_2}^2-m_{\mu}^2+m_{N_j}^2)+x(m_{W_2}^2+
m_{\mu}^2-m_{N_j}^2)-m_{W_2}^2,\eqno(A.16)$$
$$q_x^2=x^2[m_{\tau}^2y^2+y(m_{S_1}^2-m_{\mu}^2-m_{\tau}^2)+m_{\mu}^2].\eqno(A.17)$$


\begin{thebibliography}{xxxx}
\bibitem{VK15}V. Khachatryan {\it{et al.}} [CMS Collaboration], Phys. Lett.
B {\bf{749}}, 337 (2015) [arXiv:1502.07400 [hep-ex]].
\bibitem{Gaa}G. Aad {\it{et al.}} [ATLAS Collaboration], arXiv:1508.03372
[hep-ex].
\bibitem{GA17}G. Aad {\it{et al.}} (ATLAS), Eur. Phys. J. C {\bf{77}}
(2017) 70, arXiv:1604.07730 [hep-ex].
\bibitem{VKH16}[71]V.Khachatryan {\it{et al.}} (CMS), Phys. Lett. B
{\bf{763}} (2016) 472, arXiv:1607.03561 [hep-ex].
\bibitem{CMS17}CMS Collaboration, Search for lepton flavor violating decays
of the Higgs boson to $\mu\tau$ and $e\tau$ in proton-proton
collisions at $\sqrt{p}=13$ TeV, (2017), CMS-PAS-HIG-17-001.
\bibitem{KC16} K.Cheung, W.Y.Keung and P.Y.Tseng, Phys. Rev. D
{\bf{93}} (2016) 015010.
\bibitem{SB16} S.Baek and K.Nishiwaki, Phys. Rev. D {\bf{93}} (2016)
015002.
\bibitem{Eah16} E. Arganda {\it{et al.}}, Phys. Rev. D {\bf{93}} (2016)
055010.
\bibitem{AA16} A. Abada {\it{et al.}}, JHEP {\bf{1602}} (2016)
083.
\bibitem{MDCa}M.D.Campos, A.E.C.Hernandez, H.Pas, and E.
Schumacher, arXiv:1408.1652.
\bibitem{ACri}A.Crivellin, G.D'Ambrosio, and J. Heeck, arXiv:1501.00993.
\bibitem{JLD00}J.L.Diaz-Cruz and J.J.Toscano, Phys. Rev. D
{\bf{62}} (2000) 116005.
\bibitem{AAD14}A.Abada, M.E.Krauss, W.Porod, F.Staub, A.Vicente and C.
Weiland, JHEP {\bf{1411}} (2014) 048.
\bibitem{ICP74}J.C.Pati and A.Salam, Phys. Rev. D {\bf{10}} (1974)
275.
\bibitem{RNM75}R.N.Mohapatra and J.C.Pati, Phys. Rev. D {\bf{11}} (1975) 566.
\bibitem{GSRN}G.Senjanovic and R.N.Mohapatra, Phys. Rev. D {\bf{12}}, (1975)
1502.
\bibitem{Bom97}G.G.Boyarkina, O.M.Boyarkin, Physics of Atomic
Nuclei, {\bf{60}} (1997) 601.
\bibitem{Bom04}O.M.Boyarkin, G.G.Boyarkina, and T.I.Bakanova,
Phys. Rev. D {\bf{70}} (2004) 113010-1.
\bibitem{RN81}R.N.Mohapatra and G.Senjanovic, Phys. Rev. Lett.
{\bf{44}} (1980) 912.
\bibitem{RMS77}R.N.Mohapatra and D.P.Sidhu, Phys. Rev. Lett.
{\bf{38}} (1977) 667.
\bibitem{NGD91} N.G.Deshpande, J.F.Gunion, B.Kayser, F.Olness,
Phys. Rev. {\bf{D44}}, 837 (1991).
\bibitem{Bom200}G.G.Boyarkina, O.M.Boyarkin, Eur. Phys. J. C {\bf{13}}
(2000) 99.
\bibitem{JG96}J.F. Gunion {\it{et al.}}, Production and Detection at SSC
of Higgs Bosons in Left-Right Symmetric Theories, in Proc. of the
1986 Summer Study on the Physics of the Superconducting
Supercollider, June 23-July 11, 1986,
http://inspirehep.net/record/20858.
\bibitem{FA15}A.Falkowski {\it{al.}}, High Energy Phys. {\bf{57}}
(2015) 2015,  arXiv:1502.01361 [hep-ph].
\bibitem{SIG16}S.I Godunov {\it{al.}}, Eur. Phys. J. C {\bf{76}}
(2016) 1, arXiv:1503.01618 [hep-ph].
\bibitem{OMB14}O.M.Boyarkin, G.G.Boyarkina, Phys. Rev. D
{\bf{90}} (2014) 025001.
\bibitem{CP16}C. Patrignani {\it{et al.}} (Particle Data Group),
Chin. Phys. C, {\bf{40}}, (2016) 100001.
\bibitem{ATC17}ATLAS Collaboration Phys.Rev. D {\bf{96}} (2017)
052004 [arXiv:1703.09127].
\bibitem{PSB13}P.S.Bhupal Dev, Chang-Hun Lee, R.N.Mohapatra, Phys.
Rev. D {\bf{88}} (2013) 093010.
\bibitem{SD16}S.Dell'Oro {\it{et al.}}, Adv. High Energy Phys.
{\bf{2016}} (2016) 2162659, arXiv:1601.07512 [hep-ph].
\bibitem{JB13}J. Barea, J. Kotila, and F. Iachello, Phys. Rev. C
{\bf{87}} (2013) 014315.
\bibitem{CC13}Chien-Yi Chen, P.S.Bhupal Dev, and R.N.Mohapatra,
Phys.Rev. D {\bf{88}} (2013) 033014.
\bibitem{TGR82}T.G.Rizzo, Phys.Rev. D, {\bf{25}} (1982) 1355.
\bibitem{ATLAS1}The ATLAS collaboration, Projections for measurements of
Higgs boson cross sections, branching ratios and coupling
parameters with the ATLAS detector at a HL-LHC, Tech. Rep.
ATL-PHYS-PUB-2013-014 (CERN, 2013)
\bibitem{CMS1}The CMS Collaboration (CMS), Proceedings, Community Summer
Study 2013: Snowmass on the Mississippi (CSS2013) (2013),
arXiv:1307.7135 [hep-ex]
\bibitem{BBF13}H. Baer {\it{et al.}}, (2013), arXiv:1306.6352
[hep-ph].
\bibitem{TLEP}M. Bicer {\it{et al.}}, JHEP {\bf{01}} (2014) 164,
arXiv:1308.6176 [hep-ex]
\end{thebibliography}
\end{document}